\documentclass{elsart}

\usepackage{graphicx,amssymb}
\usepackage{epsfig}

\def\bfi{\begin{figure}}
\def\efi{\end{figure}}
\def\bc{\begin{center}}
\def\ec{\end{center}}

\begin{document}

\begin{frontmatter}



\title{Track Extrapolation and Distribution for the CDF-II Trigger System }


\author{Robert Downing},
\author{Nathan Eddy},
\author{Lee Holloway},
\author{Mike Kasten},
\author{Hyunsoo Kim},
\author{James Kraus},
\author{Christopher Marino},
\author{Kevin Pitts}\ead{kpitts@uiuc.edu},
\author{John Strologas},
\author{Anyes Taffard}
\address{University of Illinois at Urbana-Champaign, \\ 
1110 West Green Street, Urbana, IL 61802, USA}

\begin{abstract}
The CDF-II experiment is a multipurpose detector  designed
to study a wide range of processes observed in 
the high energy proton-antiproton collisions produced by  
the Fermilab Tevatron.  With event rates greater than 1MHz,
 the CDF-II trigger system is crucial for 
selecting interesting events for subsequent analysis.
This document provides an overview of the Track Extrapolation
System (XTRP), a component of the CDF-II trigger system.  
The XTRP is a fully digital system that is utilized in
the track-based 
selection of high momentum lepton and heavy flavor
signatures. 
The design of the XTRP
system includes five different custom boards
utilizing discrete and FPGA technology 
residing in a single VME crate.  
 We describe the design, construction, commissioning
and operation of this system.

\end{abstract}

\end{frontmatter}

\section{Introduction}
\label{sec:intro}


The CDF experiment was originally proposed more than
25 years ago.  After several years of accelerator and
detector construction, the CDF detector observed the 
first Tevatron $p\overline{p}$
collisions in 1985.  In the 20+ years since the first
collisions were observed, the experiment has undergone
several upgrades and completed several successful 
data runs, including Tevatron Run I from 1992-1996, which
yielded the discovery of the top quark \cite{top}.  While the Fermilab
Main Injector was being constructed in the late 1990's,
CDF underwent a major upgrade, which included entirely
new tracking systems, front-end electronics and a totally
new trigger system.  The first Tevatron Run II 
collisions were observed with the upgraded
CDF-II detector in 2001.  This paper documents the
CDF Track Trigger Extrapolation System (XTRP), one 
of the components of the CDF-II trigger.

Central tracking has always been one of the strengths
of the CDF experiment.  The original CDF trigger included
a three level architecture that included central tracking
information at the second level.  This tracking information
was found to provide an extremely powerful handle for 
controlling trigger rates while maintaining high efficiency
for physics signatures involving charged tracks.   With 
Tevatron Run II upgrade anticipated to provide an
instantaneous luminosity more than a factor of 10 higher
than had been seen in Run I, it was clear that central
tracking would need to play an even larger role in the
CDF trigger strategy.  This push for higher luminosity,
coupled with continual developments in high speed digital
electronics, led to a CDF II trigger design that
featured high precision central tracking at the first
trigger level, operating on every $p\overline{p}$ 
collision.

In addition to finding charged tracks on every
Tevatron bunch crossing, it is necessary to process
those tracks so that trigger objects, such as 
electron, muon or heavy flavor candidates can
be identified.  This processing is performed by
the XTRP system, which consists of fully custom digital
boards residing in a single 9U VME crate.

This document describes the design, construction
and operation of the XTRP system, which has been
operating as a central
component in the CDF II trigger system during 
Tevatron Run II.
In the next section, we begin by providing 
an overview of the Tevatron, the CDF experiment and
trigger system.
In Section~\ref{sec:XtrpDesign}, we present the XTRP 
specifications
and design concepts, 
as well as a detailed description
of the components of the XTRP system.  In Section~\ref{sec:XtrpTesting},
we describe the tools that were developed to test and
commission this pipelined digital system.  In 
Section~\ref{sec:Operations}, we describe the 
operation and performance of the XTRP system.  We 
conclude in Section~\ref{sec:conclusion}.

\section{The Tevatron and CDF Experiment}
\label{sec:background}


The CDF-II detector is 
designed to study a wide variety of high energy processes
produced by the Fermilab Tevatron proton-antiproton 
($p\overline{p}$) collider.   With a bunch crossing 
rate of $1.7\, \rm MHz$ and a $p\overline{p}$ collision rate
as high as $10\, \rm MHz$, it is necessary to perform significant
event reduction.  Less than $1$ in $10,000$ ($0.01\%$) events
can be stored for subsequent analysis, so
a fast, efficient, intelligent trigger system is crucial
to fully exploit the wealth of physics probed by
these collisions.

This section provides information on the Fermilab Tevatron,
as well as the CDF detector and trigger system.  We attempt
to present information relevant to help the reader understand
the context in which the XTRP system was designed and 
operates.

\subsection{The Fermilab Tevatron}

The Fermilab Tevatron is a high-energy, high-rate
($p\overline{p}$) collider that
permits the study of the structure of matter at very small distance
and very high mass scales.  The CDF-II detector is a 
multi-purpose experiment
specifically
designed to study $p\overline{p}$ collisions at the Tevatron.  
The CDF-II detector utilizes
precision tracking and calorimetry to identify and measure
leptons, jets and heavy flavor ($c$ and $b$ quarks).  

The first Tevatron/CDF physics run took place in 1987.  Since then, both
the accelerator and the experiment have undergone multiple rounds of
upgrades.  After collecting $110\, \rm pb^{-1}$
of data in Run I, the
Tevatron underwent a major upgrade which included the construction of
the Main Injector and Antiproton
Recycler \cite{mi}.  During the same period, the CDF detector
underwent a major upgrade to handle higher instantaneous luminosity and
higher collision rates provided by the upgraded Tevatron \cite{tdr}.

Tevatron Run II began in March 2001 and will continue throughout much
of this decade.  In the Run II configuration, the Tevatron
collides 36 proton bunches with 36 antiproton bunches with a bunch
spacing of $396\, \rm ns$.  The bunch structure of the Tevatron 
has 3 bunch trains of 12 bunches each.  Between the bunch trains are
``abort gaps'' of $3.5\, \mu\rm s$ each, and the actual bunch crossing 
rate is $1.7$MHz.  The  $p$ and 
$\overline{p}$ beam energies
are $980\, \rm GeV$, yielding a $p\overline{p}$ center of mass collision
energy of $1960\, \rm GeV$.   In July 2004, a peak luminosity of $1\times
10^{32} \rm cm^{-2} s^{-1}$ was achieved.  At this luminosity, there
are an average of 3 $p\overline{p}$ interactions per beam crossing.
The ultimate instantaneous
luminosity is expected to be $ 3\times
10^{32} \rm cm^{-2} s^{-1}$, which corresponds to an average
of  $8$ $p\overline{p}$ interactions
per beam crossing.

\subsection{The CDF-II Detector}

The CDF-II detector  has a cylindrical geometry, and
therefore utilizes a cylindrical coordinate system, with
the $z$-axis pointing in the
direction of the proton beam, $r$ the perpendicular 
distance from the
beamline and $\phi$ the azimuthal angle.  It is also 
convenient to define $\theta$ as the polar angle relative
to the $z$ axis, from which we use pseudorapidity,
which is $\eta = -\ln[\tan(\theta/2)]$.  

The
Central Outer Tracker (COT) is a open-cell drift chamber 
that provides charged track identification in the central
region ($|\eta| \le 1.1$) for tracks
with transverse momentum $p_T> 400\, \rm MeV/{\it c}$ \cite{cot}.  
The COT has
a cylindrical geometry of eight alternating axial and stereo 
superlayers.  The active volume of the COT covers
$|z|<155\, \rm cm$ and $40$ to $140\, \rm cm$ radius. 
The COT
sits within a superconducting solenoid that provides
a $1.4\, \rm T$ axial magnetic field.

Tracks found in the COT
are extrapolated inward and matched to hits in the silicon
microvertex detector for heavy flavor identification \cite{svx}.
The microvertex detector provides precise three-dimensional
track reconstruction, which is important for identification
of long-lived particles.

Between the COT and solenoid is a time-of-flight system (TOF) 
for particle identification.
Outside the solenoid are electromagnetic and hadronic 
calorimeters arranged in a projective-tower geometry, 
covering the pseudo-rapidity region 
$|\eta|<3.5$.  The central calorimeter consists of 
24 ``wedges'' in azimuth, each wedge covering $15^\circ$.
At a
depth of $\sim \! 6$ radiation lengths (shower maximum) within
the electromagnetic calorimeters are wire chambers 
to precisely measure the shower position.   An electron
is identified as a track in the COT matched to a 
cluster in the electromagnetic calorimeter (with little
or no hadronic energy) and shower
maximum position consistent with the extrapolated 
track.

Outside the hadron calorimeters are drift chambers
and scintillator counters for muon identification.
The muon systems are segmented into four components,
the Central Muon system (CMU) provides coverage
for $|\eta|<0.6$ and $p_T > 1.3 \rm \, GeV/{\it c}$.
The CMU follows the same $24$-fold azimuthal symmetry
as the central calorimeter.
The Central Muon upgrade (CMP) covers the same 
pseudorapidity region as the CMU, but sits behind
an additional $\sim \! 3$ interaction lengths 
of material, providing 
identification for muons with $p_T > 3 \, \rm GeV/{\it c}$,
with higher purity than muons identified only in the CMU.
The Central Muon extension (CMX) provides coverage
for $0.6<|\eta|<1.1$ and $p_T > 2.0 \rm \, GeV/{\it c}$.
The Intermediate Muon system (IMU) provides coverage
for $1.1<|\eta|<2.0$ and $p_T > 3.0 \rm \, GeV/{\it c}$.
Muons are identified as tracks in the COT matched to 
track segments in one or more of the muon systems.

\subsection{The CDF-II  Trigger System}

The CDF-II data acquisition system can store data
at a maximum rate of $18\, \rm MB/s$.  With an average
event size of $170\, \rm kB$, this translates into
an event rate of $100\rm \, Hz$.  Therefore, in 
processing the $1.7\, \rm MHz$ of collision data, the
CDF-II trigger system must reject more than $99.99\%$ of
the events.  In
order to maintain high efficiency for interesting signatures, the
trigger system must be fast, flexible and operate with low deadtime.
Central tracking plays a crucial role in the CDF-II trigger, allowing
for the identification of high $p_T$ electrons,
muons, tau leptons, and daughter tracks from heavy flavor decays.

The CDF-II trigger system has a three-level architecture
with each level providing a rate reduction sufficient to allow for
processing at the next level with minimal deadtime.  Level-1 operates
on every beam crossing and uses
custom designed hardware to find physics objects based on a subset
of the detector information and makes a decision based on simple
counting of these objects (\it e.g. \rm one $12\, {\rm GeV}/c$ 
electron or two
$1.5\, {\rm GeV}/c$  muons). The Level-2 trigger uses custom hardware to do a
limited event reconstruction which can be performed in
a single programmable processor.  
The Level-3 trigger uses data from 
full
detector readout to fully reconstruct events in a farm 
consisting of more than 100 commercial dual processor
PCs.

The functionality of the three level pipelined and buffered
trigger system is shown in Figure~\ref{fig:dataflow}.
Although the time between collisions is $396\, \rm ns$, the
base CDF-II
clock period utilized by the entire CDF trigger
system (CDF\_CLOCK) is $132\, \rm ns$.
The original
specification for the CDF-II detector allowed for operation
with a Tevatron bunch spacing of $132 \, \rm ns$, and 
consequently  the
CDF-II trigger and data acquisition system was built to
operate in both $132$ and $396\, \rm ns$ configurations
from a base system clock period of $132\, \rm ns$.  
To allow time for transmission and processing of the trigger
signals, there is a 5.5 $\mu$sec
Level-1 latency from $p\overline{p}$ collision to Level-1 trigger
decision.  This requires each detector element to have
local data buffering for 42 clock cycles.

If an event is accepted by the Level-1 trigger, all front-end
readout 
components move the data to one of four on-board Level-2 buffers.
This buffering is sufficient 
to average out the rate fluctuations and allow
a 25 kHz Level-1 accept rate with $\le 5\%$\ deadtime for the
average of 35 $\mu$sec Level-2 processing time.  An event which
satisfies the Level-2 trigger undergoes full detector readout,
and the data from each of the front-end elements is assembled into
a single event which is fed to one of the processors in the 
Level-3 computing farm.   Detector readout,
the event builder and Level-3 computing farm provide
sufficient bandwidth to permit the 
Level-2 trigger to accept events at a rate of 
$350$ Hz.  In the Level-3 trigger processor farm, 
the events are reconstructed
and filtered using full event reconstruction, with $75$ Hz
written to permanent storage.  The peak achieved 
rates are $15$-$25\%$ higher than the typical 
operating trigger rates listed here.

The block diagram for the CDF~II trigger system is presented in
Fig.~\ref{fig:trigger}.  The input to the Level-1 hardware comes
from the calorimeters, tracking chamber, and muon detectors. The
decision to retain an event for further processing is based on the
number and energies of track, electron, photon,
muon, $\tau$ lepton and jet candidates,
as well
as the total energy and missing transverse energy 
in the event.  The Trigger Supervisor System (TSI) 
is responsible
for maintaining synchronization and allocating buffer 
space for each event accepted at Level-1 \cite{tdr}.

Since the Level-1 and Level-2 trigger systems each 
require rejection
factors of $\sim \! 70$, they need
significant detector information to perform their function.  
Several detector systems provide information for
the Level-1 trigger 
decision:  the calorimeter (CAL), COT, 
CMU/CMX/IMU systems (MUON), TOF,  
Cerenkov luminosity
monitor and Roman Pot detectors.

Two components add additional information in the Level-2
trigger system, the Silicon Vertex Trigger (SVT) and the
shower maximum wire chambers (CES).  
The SVT \cite{svt} 
incorporates
information from  the high precision silicon 
microvertex detector \cite{svx}
into
tracking trigger selection. The SVT provides, for the
first time in a hadron-collider experiment, the ability to 
trigger on
displaced tracks arising from the decay of long-lived particles.
This has already produced a number of new
results involving hadronic decays of $c$ and $b$-quarks \cite{hf}.

All of the information available in the Level-1 trigger system
in addition to data from the SVT and CES are brought together
in the Level-2 decision crate (GLOBAL L2).  The Level-2 
decision node is 
the first place where software algorithms are utilized to 
process the event.  
For the first part of Run~II,
a DEC Alpha processor mounted on a custom
9U VME board was the Level~2 decision node.  
As part of a Level-2 trigger upgrade, the Alpha was
replaced by a commercial PC processor utilizing a PCI interface.  
The Level-2 decision node is pipelined
so that one event may be processed while the next event
is being loaded.

\bfi[htbp]
\bc
\includegraphics[width=14cm]{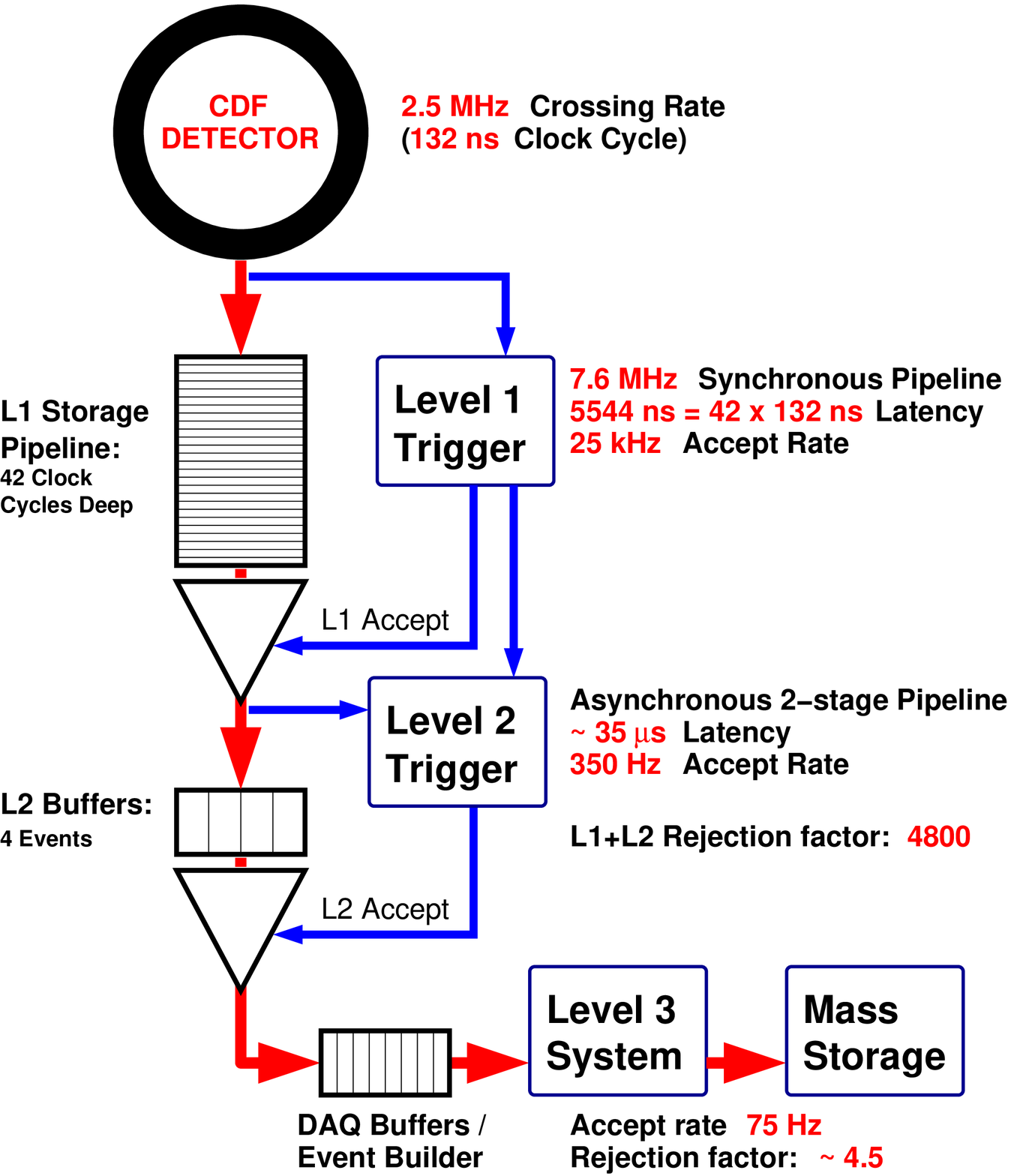}
\ec
\caption{The CDF-II trigger and data acquisition system. Data is
acquired at the beam crossing period of $396\, \rm ns$ into a
synchronous pipeline that is clocked at $132\, \rm ns$.
The Level-1 decision is produced after 42
clock cycles, at which point event processing becomes
asynchronous.  Typical trigger rates and rejection factors for
the three-level system are shown in the figure.} \label{fig:dataflow} \efi

\bfi[htbp]
\bc
\includegraphics*[width=12cm]{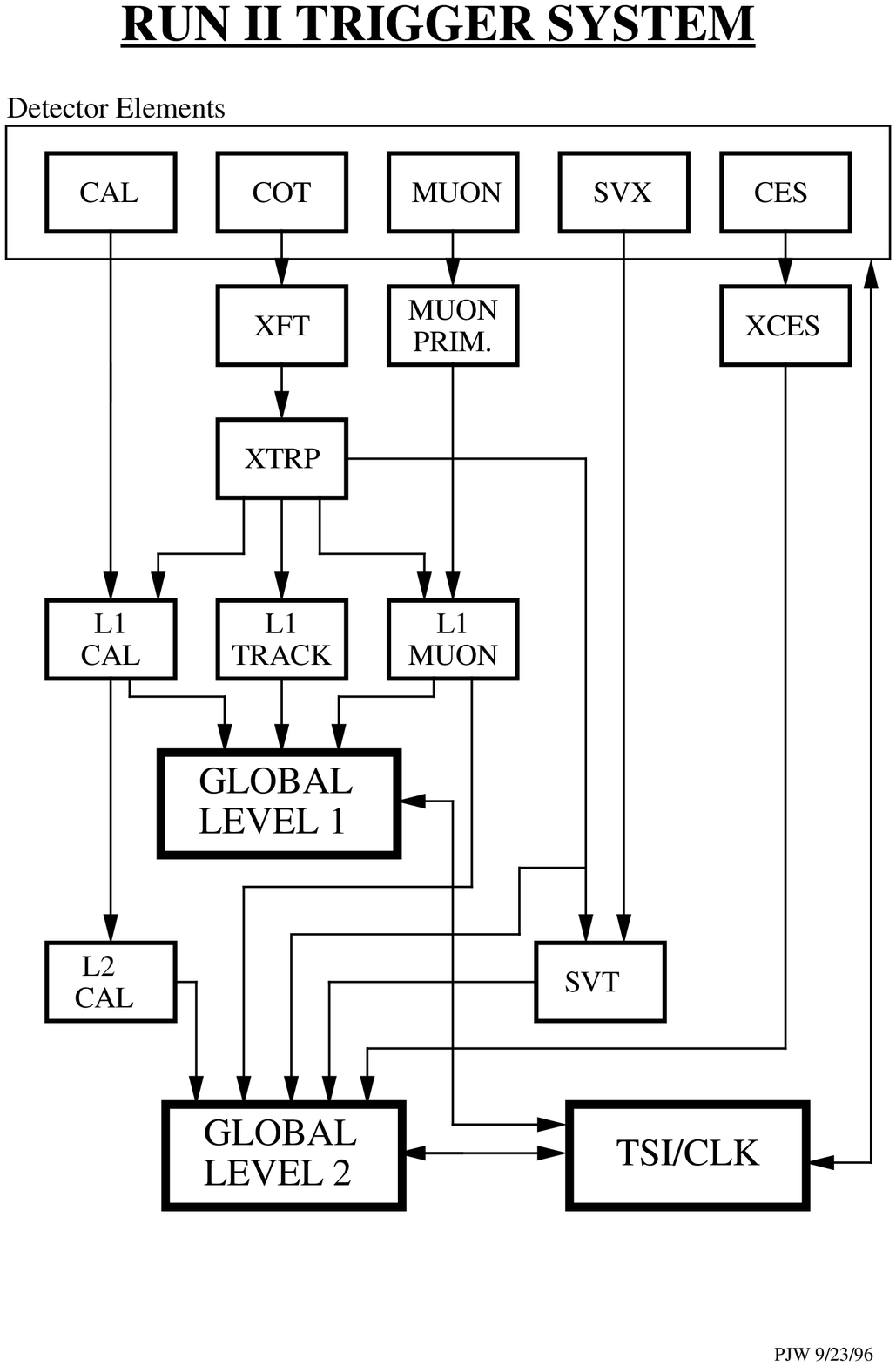}
\ec \caption{The CDF-II trigger system.  Trigger primitives are
acquired from the detector elements and lead to a Level-1
decision.  In the case of the track-based triggers, the XFT finds
tracks in the COT, which are passed to the XTRP system.
In the XTRP, the tracks are extrapolated to the muon and calorimeter
systems for muon and electron identification.  Tracks are also passed
onto the Track
Trigger (L1 Track) by way of the XTRP.  The XTRP additionally provides
tracking information for the Silicon Vertex Trigger (SVT) and
Level-2 trigger processor.} \label{fig:trigger} \efi

\subsection{The XFT System}

Track-based triggers (including lepton and heavy flavor triggers),
account for approximately
$75\% $ of the CDF-II trigger bandwidth.
The XFT system utilizes hit information from the
COT to perform charged track reconstruction 
in the $r$-$\phi$ plane at Level-1.  For 
tracks with $p_T >1.5\, \rm
GeV/{\it c}$, the XFT system has high
efficiency ($>90\%$), good transverse 
momentum resolution, $\delta p_T/p_T =0.002p_T$,
and 
pointing resolution, $\delta \phi_0 = 0.002\, \rm radians$,
where $\phi_0$ is the azimuthal angle of the track 
measured at the beamline ($r=0$.) \cite{xft}.

The XFT logically divides the COT into 288 azimuthal segments,
each covering $1.25^\circ$.  These segments are processed in
$24$ XFT ``Linker'' boards, with each board covering $15^\circ$ in
azimuth.  The segmentation is well-matched to the XFT 
angular resolution and the symmetry of the central
calorimeter and CMU.
The XFT reports no more than one track per 
segment.  If more than one track is identified within
a single $1.25^\circ$ segment, the track with highest
transverse momentum is reported.

Tracks found by the XFT are passed to the XTRP 
system, where the tracks are extrapolated to the
calorimeter (L1 CAL) and muon systems (L1 MUON) 
for lepton identification.  The
XTRP also passes tracks to the Track 
Trigger (L1 TRACK), which
selects events based upon trigger topology.
If an event is accepted by the Level-1 trigger,
the XTRP passes a complete list of tracks to the 
SVT and Global Level~2 systems.
\section{The CDF Extrapolation (XTRP) System}
\label{sec:XtrpDesign}

In this section, we will describe the specifications
for the XTRP system, the design and implementation of the 
system, the components and their interconnects.  
We begin with an overview of the system and its functionality,
followed by a detailed description of each of the 
components of the system.

\subsection{XTRP Overview}

The purpose of the XTRP is to receive tracking information
from the  XFT and
distribute the tracks and information derived from the tracks
to the Level-1 and Level-2 trigger subsystems.
After receiving the tracks from the XFT, signals are sent to the
Level-1 Muon system (L1 MUON),
the Level-1 Calorimeter trigger (L1 CAL), and the
Level-1 Track Trigger (L1 TRACK) as shown in Fig.~\ref{fig:trigger}.
The tracks are also put into a storage 
pipeline and upon receiving a Level-1
accept are
sent to the SVT and the
Level-2 decision processor.

The XTRP system consists of a Clock/Control Board, twelve Data
Boards, a Track Trigger Board, and two types of transition
modules. 
The entire XTRP system resides in a single, 9U
VME crate with a custom J3 backplane that satisfies
the VIPA specifications \cite{vipa}. The crate also
contains a commercial VME CPU \cite{cpu} 
and TRACER module which are common to
all CDF-II VME crates.  The TRACER (TRigger And Clock + Event
Readout) is the gateway between the XTRP
crate and the CDF-II trigger system. 
It receives CDF-specific timing signals
as well as Level-1 and Level-2 Trigger Decisions  \cite{tdr}.

For each Tevatron bunch crossing, 
the XFT reports information for all 288 azimuthal 
track segments to
the XTRP.  For the majority of the segments, the XFT 
reports that
no track was found.  Reporting a fixed amount of information 
on each bunch crossing lends itself nicely to the
synchronous data processing performed in the Level-1
trigger system.  Track finding in the XFT is complete
$2.7\, \mu$sec after each $p\bar{p}$ collision \cite{xft}.
All of the XFT data is transferred to the XTRP in
$132\, \rm ns$.

The XTRP receives the track data from the XFT and,
through the use of lookup tables,
calculates the relevant information
required by other systems to construct trigger objects.
For example, muon primitives (track segments in
the central muon chambers) are found at the same time the XFT is
finding tracks in the COT.  The XFT tracks are sent to the XTRP,
which informs the Level-1 Muon trigger of all locations where
a track extrapolates to the central muon systems.
The definition of a muon
object
in the trigger is a track in the central muon system  that is consistent
with an extrapolated central track.  Similarly, an electron is
defined as a track plus an electromagnetic
shower, with the XTRP extrapolating the tracks into the calorimeter.

The following information is sent to the Level-1 trigger subsystems
from the XTRP:
\begin{itemize}
\item {\bf Central  Muon systems (L1 MUON).} XFT tracks are extrapolated
to the radii of the CMU, CMX and IMU.
One
or more bits, corresponding to $2.5^\circ$ azimuthal 
segmentation, are set according
to $p_T$, $\phi $, and amount of multiple scattering.  These bits are sent
to the Level-1 Muon Trigger system. Two separate $p_T$ thresholds are
available for each of the three (CMU, CMX, IMU) subsystems.

\item{\bf Central Calorimetry (L1 CAL).}  XFT tracks are extrapolated
  to Central Calorimeter towers.
A set of four bits for each $15^\circ$
wedge is sent to the Central Calorimetry Level-1 trigger.  These
bits correspond to four separate momentum thresholds.

\item {\bf Level-1 Track Trigger (L1 TRACK).} The Level-1 Track Trigger
is an adjunct
to the XTRP.  It resides in
the same VME crate and  provides Level-1 triggers based on XFT track
information only.  The XTRP modules select tracks above a given $p_T$
threshold and passes them on a bus to the Track Trigger. The total number of
tracks  is counted.  If more than 6   tracks
 are found an automatic Level-1 accept is generated.  If there are 6
 tracks or fewer, the $p_T$ and $\phi $ information is used to interrogate
look-up tables to generate up to 
15 distinct Level-1 track-only triggers.
\end{itemize}
The XTRP must provide 
output information to the L1 CAL,  L1 MUON and 
L1 TRACK systems within 
$300\, \rm ns$ of having received input XFT data.
The L1 TRACK decisions must be available $396\, \rm ns$ 
after having received all of the input track data.

Upon receiving input data from the XFT, 
all segments  are put into a pipeline 
and stored pending the Level-1 trigger
decision.  If a Level-1 accept is received
the tracks are latched into Level-2 buffers.
All non-trivial tracks
are then extracted and put into two separate FIFO's for delivery to the
Level-2 processor and to the SVT respectively.

An overview of the XTRP system can be seen in Fig.~\ref{fig:xtrpData}.
In the following subsections, we provide a detailed
description of the XTRP system.

\bfi[htbp]
\bc
\includegraphics[width=14cm]{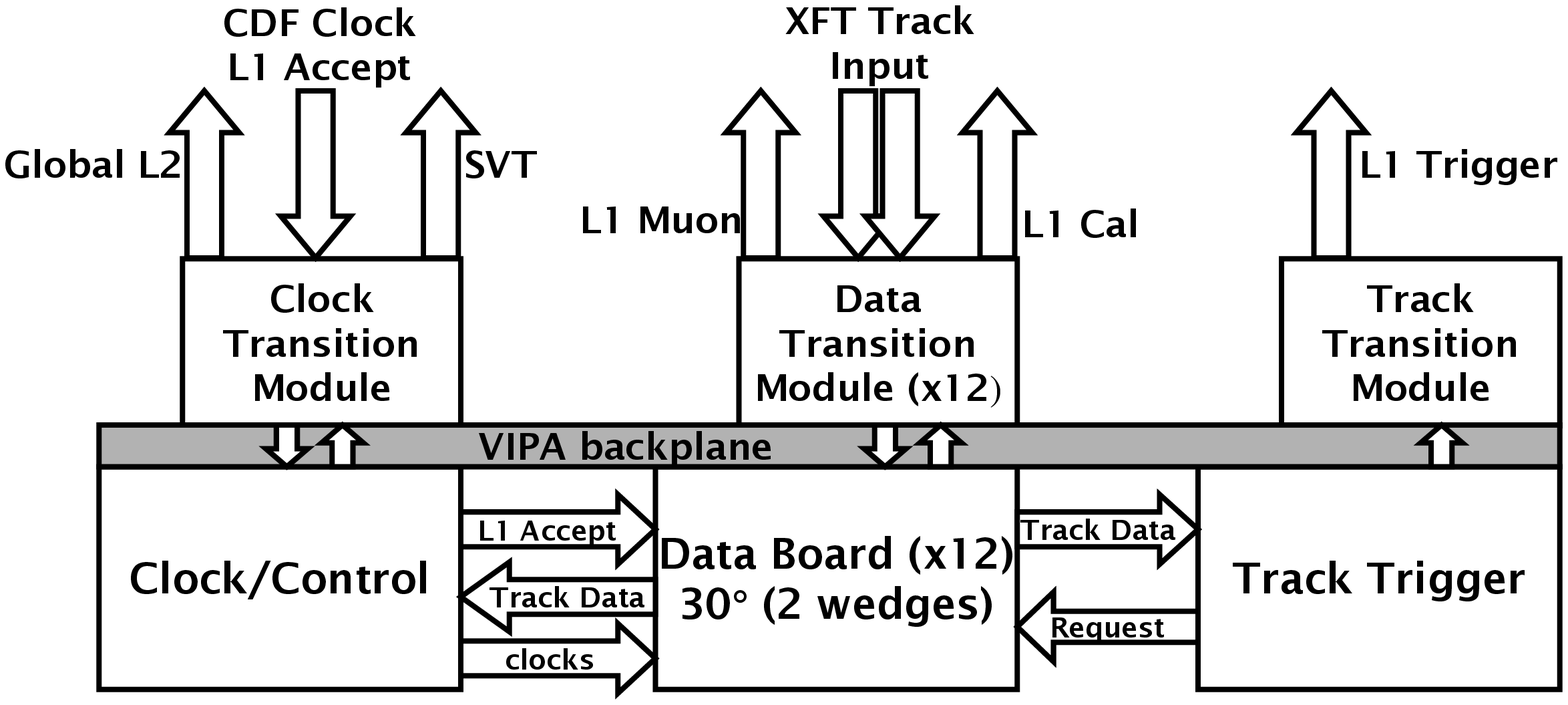}
\ec
\caption{An overview of the XTRP system.  Each 9U VME board 
(Clock/Control, 12 Data Boards, Track Trigger) has an associated
transition module.  Horizontal arrows between the boards indicate
data transferred within the system via the VIPA backplane.  Vertical
arrows within the backplane indicate pass-through I/O between the
board and its transition module.  Vertical arrows at the 
top of the picture indicate cables connections which are 
data interfaces with other trigger systems.  All cables 
connect to the system through transition modules as shown
in the figure.} \label{fig:xtrpData}
\efi

\label{subsec:SysDesignOverview}

\subsection{Clock Control Board}
\label{subsec:ClockBd}

    The Clock Control Board has two primary functions:
distribution of CDF-specific signals to the Data Boards and
Track Trigger and interface with the Level-2 processor and the SVT.

The  Clock/Control board receives the timing
signals, including the $132\rm~ns$ CDF\_Clock, and provides both
$132\rm~ns$ and $33\rm~ns$ clock signals
to the Data Boards and Track
Trigger Board.  The Clock/Control board can 
delay the output clock signals up to
one full cycle in $3\rm~ns$ increments.  These
variable delays are 
used to synchronize
the XTRP system to the phase of the 
incoming track information
from XFT.  A fully programmable delay was implemented to
account for uncertainty in the cable delays on the incoming
track data.

    Although a custom VME backplane was developed for CDF-II VME
usage, the XTRP system utilizes a standard VIPA backplane.
In a CDF-II VME backplane, signals are bussed across the P2 connectors,
and may have clock skew of $\sim \! 7 \rm \, ns$ from one end of
the crate to the other.
The  VIPA backplane does not carry any common signals across the
P2 connectors.
In order for the Clock Control Module to receive the CDF-specific
signals from the TRACER, a custom P2 transition card is installed
opposite the TRACER.  The signals are carried via connector to the
Clock Control transition module and then fed through the
backplane to the Clock Control module.

    The $132\rm~ns$ PECL CDF Clock signal is converted to TTL and passed
through a programmable tap filter and Roboclock chip.  This produces
an even duty-cycle $132\rm~ns$ clock signal. Additionally,
a $16.5\rm~ns$ clock signal is also generated to
facilitate both the coarse delay of the $132\rm~ns$ clock
and generate an even duty-cycle $33\rm~ns$ clock. The
refined $132\rm~ns$ and $33\rm~ns$ clocks
are multiplexed against signals,
regulated by VME-access, that act as clocks. The effect is
that the XTRP system can run at real-time speed governed by
CDF\_Clock or run as a large state machine, with each step
interval under user control.  The latter mode allows the
user to ``step'' data incrementally through the
system, while monitoring the data through each stage. This 
is described in Section~\ref{sec:XtrpTesting}.

    The multiplexed clock signals are fanned-out on the
Clock Control Board and fed to the custom backplane where
each of the Data Boards and the Track Trigger 
receive a unique set
of clock signals. With trace length matching, the skew for
these signals less than $1\rm~ns$ across all of the boards.

    For every Level-1 accept the Clock Board must send a
list of tracks to the Level-2 and SVT systems.  This track
list is created via a token
ring initiated by the Clock Control board.  The token ring
is carried from board-to-board via the J5 connector on 
the custom J3 backplane.
The Clock Board issues the token, which traverses 
the Data Boards and returns to the Clock Board. As the
token arrives at each Data Board FPGA which holds the track 
data on the Data Board, 
the FPGA sends the track data across
the custom backplane to the Clock Board. The Clock Board
has two parallel on-board discrete FIFOs used to accumulate 
the track data.  The FIFOs contain identical information, one
sending the output to the Level-2 trigger and the other 
sending output to the SVT system.  Data transmission is
differential LVDS via ribbon cable.  Since
these transfers are variable length and  
asynchronous, it is terminated by an end-of-event
word.  Additionally, an 8-bit ``bunch counter'' is 
appended to uniquely identify the event.  The value
of this counter is compared with the expected value
on the receiving end to verify that all parts of the
system are processing the same event.


\subsection{XTRP Track Input}
\label{sec-input}

On each event, track data for each wedge
are sent from XFT to the XTRP system within a single
$132\rm~ns$ clock cycle.  As mentioned previously, the
XFT reports information for each segment 
regardless of whether a track is
found or not.  For each segment,  13 bits of track
information are reported:
\begin{itemize}
 \item    {\bf 7 bits of transverse
momentum ($p_T$) information:}
The XFT finds tracks with $p_T>1.5\, {\rm GeV}/c$.  
Track $p_T$ values reported by the XFT are 
encoded into 96 bins, with
bins 0-47 corresponding to negatively charged tracks
and bins 48-95 corresponding to positively charged tracks.
If no track is found in a given segment, its momentum
bits are assigned a value of 124.   The ``short'' tracks
have a poorer $p_T$ resolution, hence the $p_T$ bin definitions
are different for short tracks.
\item {\bf 3 bits of ``local $\phi$'':}  Each $1.25^\circ$ segment
is passed from the XFT to XTRP on a specific data bus, so the 
location of a segment in
increments of $1.25^\circ$ is known by hardware location.  In
addition, three bits determine the azimuthal position of
the track within the segment to an accuracy of 
$\sim 0.16^\circ = 1.25^\circ/8$.   The reported 
azimuthal position of the track reflects the position of the track
at a radius of $130\, \rm cm$ from the beamline.  
\item {\bf 1 isolation bit:}  This bit was intended 
to indicate whether or not there were
other tracks nearby in $\phi $.  To date, this bit is unused.
\item {\bf 1 ``short track'':}  This indicates that a found track did
not pass through the outermost layer of the COT, denoting a track
at larger pseudo-rapidity ($|\eta|>1$).
\item {\bf 1 undefined bit:}  This bit is reserved for future use.
\end{itemize}

These 13 bits of information are shipped from the XFT to the XTRP
for all 288 segments for each event.  
The data are  sent over 24 (one per wedge) 100-pin
differential signal cables. 
The signal cables are bundled twisted pair, with a
transmission length of 10 meters, terminated with
subminiature D-style connectors.  The data transmission
is by differential LVDS format, synchronous with
the $132\, \rm ns$ CDF clock.
In addition to the segment
information listed above, a ``bunch 0'' bit and 
8-bits of  bunch crossing
number are sent on every cable to verify synchronization. 
To accomplish 
the transfer of an entire event  every $132\rm~ns$,
the data are 1-to-4 multiplexed and sent with a $33\rm~ns$
clock.  For a single event, 24 bytes are transferred per
cable (corresponding to $15^\circ$ in azimuth).  Integrated
over 24 cables, this translates
into a data rate above $4 \, \rm GB/s$ for the entire
system.

The XFT data are received by the
Data Board Transition Modules, which receive the differential
LVDS signals and translate them to TTL before passing them
through the P2 connector to the Data Board. 
There are eight Xilinx FPGAs on each Data Board 
which receive track information
for each wedge.  They are referred to as ``Pipe'' FPGAs because they
decode and pipeline the incoming track data.  Each Pipe FPGA handles
track data for 3 XFT segments, so there are four Pipe FPGAs per
wedge.
The incoming track data for a single wedge are bussed to all four
Pipes and each Pipe picks off the data it needs during each phase.
Every $132\rm~ns$ each Pipe assembles track data for three segments 
which are
then ready to be processed.

\bfi[htbp]
\bc
\includegraphics[width=14cm]{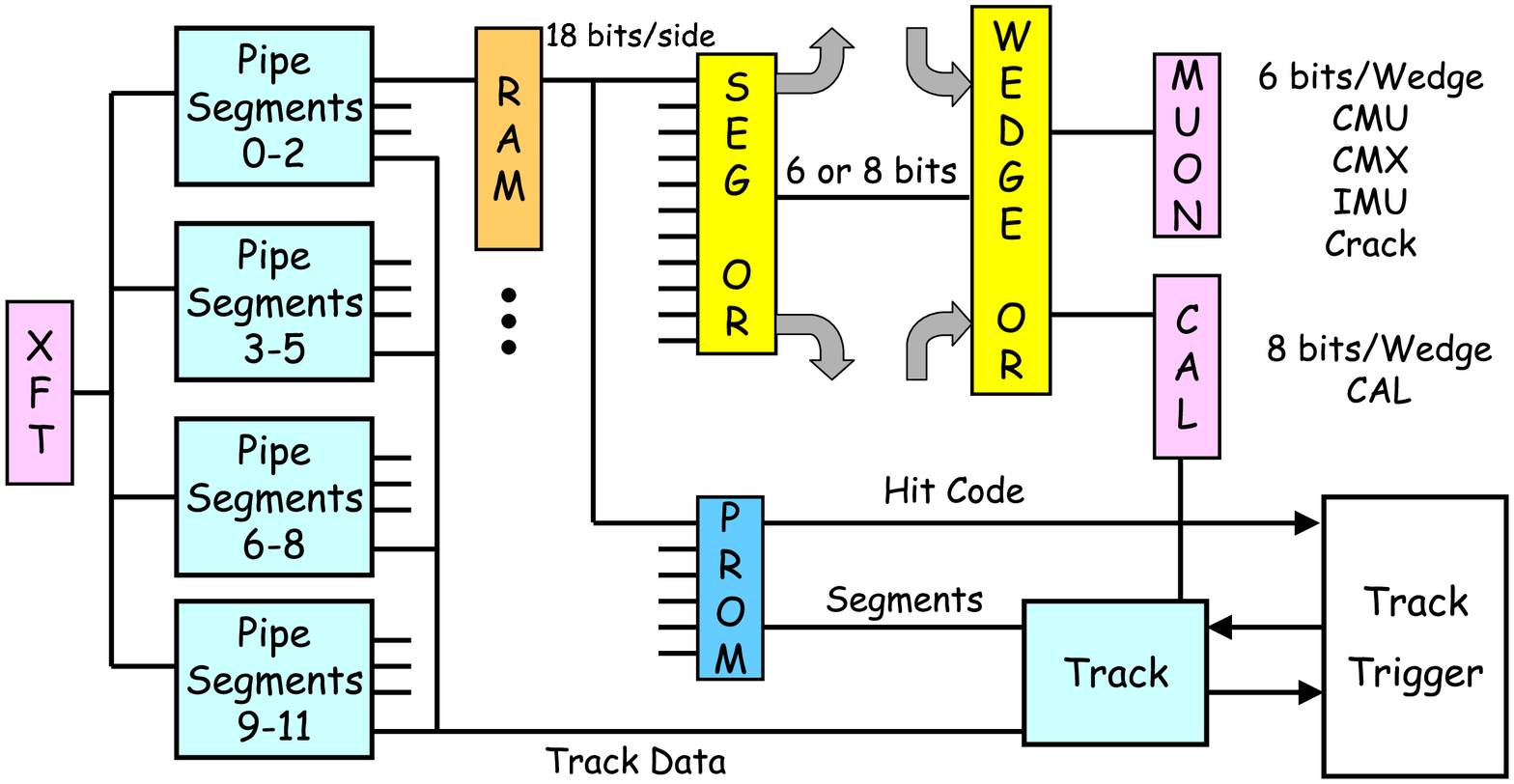}
\ec
\caption{Block diagram of the XTRP Data Board.  Input data are
demultiplexed in the Pipe FPGAs.  The data are then presented
to the lookup RAMs, followed by several stages of data
compression as described in the text.  The output stages
and Track Trigger interface are responsible for shipping
the data to other trigger systems.} \label{fig:dbblock}
\efi


\subsection{Data Boards}
\label{subsec:DataBd}


The Data Boards perform the bulk of the functionality of
the XTRP system.
Each of the twelve boards handle $30^\circ$
of the detector,   corresponding to two $15^\circ$ calorimeter
wedges.
On each clock cycle ($132\rm~ns$) the Data Boards receive
and decode incoming track data.
The
system receives  a full event 
on each $132\rm~ns$ clock-cycle.  The data
associated with non-collision clock-cycles is processed as
a normal collision event, but
trigger decisions are only relevant when there is a bunch
crossing.    Extrapolation information for muon and calorimeter
trigger systems are also generated on each clock cycle.  The
Data Board processing time between the arrival of the track data
and the production of extrapolation information is
approximately $396\rm~ns$.

An overview of the Data Board functionality is shown in
Figure~\ref{fig:dbblock}.

Data Board function begins with the
XFT track data being demultiplexed inside the Pipe
FPGAs.
The Level 1 pipeline for XFT track data are fully enclosed within the Pipe
FPGAs.  The pipeline depth is programmable up to 32 crossings 
($32\times 132\rm~ns$) after the data are received.
The track data are stored until a Level 1 Accept/Reject decision
arrives with information as to which, if any, Level 1 Decision Buffer
the track data must be stored. The Clock/Control Board  polls
Level 1 Decision Buffer data from every Pipe FPGA into a FIFO, and
sends the data to the Level-2 processor and the SVT.

\subsubsection{Clocking}

There are two distinct clock circuits on the Data Board:
Oscillator Clock and
CDF\_Clock. The Oscillator Clock circuit supplies a clock signal to all the
FPGAs to perform VME-based transactions and run state machines within the
FPGAs. The CDF\_Clock circuit provides the clocking for all the data
paths on the Data Board.

\subsubsection{Extrapolating Tracks to Calorimeter and Muon Systems}
\label{sec-extrap}

The extrapolation of each track to the muon chambers and calorimeter is
handled by lookup RAMs.  Each Data Board has 24 lookup RAMS, one for
each $1.25^\circ$ segment within the $30^\circ$ wedge covered
by the Data Board.
Each lookup RAM is $32K\times 36$ 
with 15 address bits and the 36 output bits
divided into two 18 bit outputs.
 Data pertaining to the Central Muon
chambers (CMU) and Central Muon Extension chambers (CMX) are grouped
into the ``CM side" of the RAM. Data pertaining to the Intermediate
Muon chambers (IMU), calorimetry (CAL), Track Trigger, TOF
and $\phi$-gap bits are grouped into the ``IM side" of the RAM.

The RAM contains the extrapolation
data for every possible track. Once track data has been decoded for a
segment, the 13 bits of track data along with 2 ``phase'' bits are
presented as the address to a segment RAM.  The phase bits are 
used to differentiate different lookup tables for different
subsystems or different momentum thresholds.  The data which are
stored in the RAMs are generated and downloaded before
each running period depending upon specified trigger parameters,
and is discussed in Section~\ref{subsec:interTrigTable}.
The 13 bits of track
data remain fixed for a single event ($132\rm~ns$) 
but the phase bits change
every $33\rm~ns$.  This provides four phases of output for every track.
Given the dual output and four phases, eight separate lookups  
of extrapolation
information are performed for each track, which are summarized in
Table~\ref{ta:lookup}.

For the muon lookups: CMU, CMX, and IMU, each phase corresponds
to a single $p_T$ threshold.  Each of the 18-bits of output information
correspond to a $2.5^\circ$ window in the muon system, so it is
possible for a segment in one wedge to extrapolate to the
adjacent wedges in the muon system.  The $\phi$-gap and TOF
lookups have two bits (low and high $p_T$ thresholds) for each
$15^\circ$ wedge. 
The Calorimeter lookup provides 16-bits
of information, which is 8 $p_T$ thresholds mapped to a
$30^\circ$ window in the CDF-II Calorimeter.  Since the
Calorimeter
trigger has $15^\circ$ granularity, this extrapolation
allows for tracks to extrapolate from one wedge into
the nearest neighbor wedge.
 The Track Trigger output is a single bit per wedge which
is used to determine which tracks
are eligible for transfer to the Track Trigger Board, as 
described below.

\begin{table}[htpb]
\begin{center}
\caption{Quantities extracted from the lookup RAMs. The 13-bit
track data word is presented to the RAM for $132\rm~ns$, while the
two phase bits cycle (00, 01, 10, 11 binary) each $33\rm~ns$. Each
phase represents a different lookup on each ``side'' of the RAM.
The output of each side is 18 bits of data, as described in the
text.}\label{ta:lookup} \rm
\begin{tabular}{|c|c|c|}
\hline lookup phase & CM side & IM side \\
\hline
0 & CMU high $p_T$ & Calorimeter ($8\ p_T$) + Track ($1\ p_T$) \\
1 & CMU low $p_T$ & $\phi$ gap ($2\ p_T$) + Time-of-Flight ($2\ p_T$)\\
2 & CMX high $p_T$ & IMU high  $p_T$\\
3 & CMX low  $p_T$ & IMU low  $p_T$\\
\hline
\end{tabular}
\end{center}
\end{table}





\subsubsection{Compression of Extrapolation Output}

The result of each lookup (one per segment) is $18$-data bits, corresponding
to different $p_T$ thresholds or detector $\phi$ segmentation.  Since
nearby tracks can extrapolate to the same location in the detector,
the extrapolation information must be compressed (``OR''ed) so
that the ultimate output maps exactly onto the detector geometry.
This compression is
carried out through in series of stages. Data progresses through
the stages in $33\rm~ns$ cycles. Intra-wedge compression, among adjacent
segments, is performed first, followed by inter-wedge compression,
among adjacent wedges.  The inter-wedge compression must handle
adjacent wedges across adjacent Data Boards.
The compression patterns are slightly
different between calorimetry data and muon data.  
To save space on the Data Boards, the calorimetry and muon
compression functions are performed by the same TTL/GTL
translators with open-collector outputs for wire-AND logic.

All stages within the compression region are accessible to VME,
and one can read the state of each compression stage for data
verification. Furthermore, the VME access allows one to override
the data presented to a subsequent stage with simulated data.

\subsubsection{Extrapolation Output}

After compression stages, the data are
sent by the Data Board through the backplane
to the Data Board Transition Module.
Calorimetry data for each $15^\circ$ wedge
are converted into a
differential LVDS logic level before being
sent to the Level-1 Calorimetry Trigger system \cite{tdr}
over two flat twisted pair cables, one cable per wedge.
Muon data are input to a Channel Link chip \cite{channellink}
and passed in a quasi-serial format to the
Level-1 Muon Trigger System over a single 
shielded twisted pair cable, terminated in 
an HD-20 connector.  Both the L1 CAL and 
L1 MUON outputs are synchronized to the $132\, \rm ns$
CDF\_Clock.

\subsubsection{Track Trigger Interface}

In parallel with extrapolating tracks to muon and
calorimeter systems, up to two tracks
per wedge may be sent to the Track Trigger board.
Of the tracks that pass a single programmable $p_T$ threshold,
the tracks with the largest and smallest $\phi$ within a wedge are
sent by the Data Boards to the Track Trigger board.  The
limit of two-tracks per wedge was necessitated by bandwidth
issues.  The choice of the two ``outer'' tracks (largest 
and smallest $\phi$ within a wedge) was an algorithmic 
choice that best preserves the physics targeted by this
trigger.
The Track Trigger path within
the Data Board
provides the mechanism to select and transmit the desired tracks
to the Track Trigger Board.

In the first lookup phase on the ``IM'' side, each RAM 
generates a single bit to indicate 
whether a
track above a specified $p_T$ threshold 
(typically $2\, \rm GeV/{\it c}$) 
exists in the segment track presented to the Lookup RAM.
On each Data Board the Track FPGA collects these bits and determines
whether each wedge has 0, 1 or 2 tracks eligible to be sent to the
Track Trigger. (As described above, a wedge with more than two 
eligible tracks is deemed to have two eligible tracks.)  
The Track FPGA then initiates a hand-shake with
the Track Trigger.  The information on the number of tracks per
wedge (all 24 wedges) is accumulated by the Track Trigger board.
 The Track Trigger has lookup-RAMs to
generate 3-bit codes which are returned to the Data Boards.
The Data Boards then assert track data onto the 
time-multiplexed
track data bus according to the codes generated by 
the
Track Trigger.

On each Data Board, PROMs are used to discriminate
the farthest separated tracks within a wedge that have asserted threshold
signals. One set of PROMs, Code PROMs, outputs track quantity information
based on the number of asserted threshold signals. A second set of PROMs,
Address PROMs, outputs segment identification patterns.   
The Track Trigger board
queries
the Data Board for segment data.
The Data Board sends the information for the track, in addition
to the segment location, to the Track Trigger board.


\subsection{The Track Trigger}
\label{sec-tracktrig}

The Track
Trigger board is the one board in the XTRP system that directly
renders trigger decisions.  These decisions are based upon
tracking information only.   Events selected by the track-only
path at Level-1 provide heavy flavor candidate events 
for SVT trigger at Level-2.

Details of the interface between the Data Boards and the
Track Trigger are described above.  On the Track Trigger
board, the track data are routed  to a set of six
``Sort'' FPGAs. One half of these FPGAs is dedicated to
extracting $p_T$ data ($p_T$ + isolation + short track bits),
the other half extracts the $\phi$ data.  From the
wedge and segment origin of each track, a 9-bit ``global $\phi$''
is generated ($1.25^\circ$ segmentation).  The local-$\phi$ 
information from the XFT is dropped.

The Track Trigger board receives a maximum of 6 tracks and
must evaluate every possible two-track pair for the
six tracks.   This yields ``6-pick-2'' or $15$ possible
combinations of track pairs.  Each of the the Sort FPGAs
output data (segment or $\phi$ information) for 5 pairs
of tracks.

The data are fed to a bank of lookup RAMs.  There are
15 unique $p_T$ RAMS and 15 unique $\phi$ RAMs, with
each RAM corresponding to a specific track-pair. 
Trigger selection criteria are programmed into the RAMs. 
The lookup RAMs on the Track Trigger are $512K\times 8$, with
19 address bits
and 8 data bits.  Each track provides 9 bits to the lookup RAM, 
with the $19^{th}$ bit used as a ``phase'' bit to generate
two sets of 8 trigger bits. 
For every track-pair, the $p_T$ and $\phi$ lookup 
outputs are ANDed together to generate
a trigger decision for that pair.  
All pairs are then ORed together
to generate the trigger decision 
for the event.  The RAMs
output 8 decision bits every 66ns.
The two 8-bit words are 
concatenated into a single 16-bit
trigger word.
Since one trigger bit is reserved for the
auto-accept trigger ($>6$ tracks) the Track Trigger is
capable of generating 15 different track triggers.  These
15 different triggers can be any combination of single-track
and two-track selection criteria.

The resultant trigger data is piped to a Level 2 Buffer and
fed through the backplane to the Track Trigger Transition
Module.  The trigger signals are converted to differential
LVDS and sent to the global Level-1 decision crate over
shielded twisted pair cable.  The trigger signals are
synchronized to the $132\, \rm ns$ CDF clock.

\subsection{Interface With Trigger System}
\label{subsec:interTrigTable}

Since the extrapolation parameters may change during 
different running periods, the infrastructure was 
developed so that the values loaded in the lookup
RAMs (both on the XTRP Data Boards and the Track Trigger
Board) could be generated dynamically based upon the
number and types of triggers utilized.   In this section,
we describe how the extrapolation parameters are 
generated and loaded into the XTRP system.

The Data Board lookup RAMs contain the information which
takes an input track and extrapolates it to the calorimetry
or muon systems.  The extrapolation must account for 
the geometry of the detector, track curvature from the
axial magnetic field and multiple scattering.
The extrapolation formula used is the same for all detectors.  The
extrapolation determines a ``window'' bounded by a minimum
and maximum value of 
$\phi_{detector}$ within which a particle may be found. The
extrapolation window is given by:

$$\phi_{detector} = K/p_{T} \pm \sqrt{(3\sigma_{K}/p_{T})^{2} +
\sigma_{a}^{2}} + \phi_{XFT}, $$

where the $\pm$ values yield the maximum and minimum 
values of $\phi_{detector}$ for a track that 
has a signed transverse momentum  $p_T$.
The terms $K$, $\sigma_{K}$ and $\sigma_a$ are 
constants that
depend detector subsystem.   The term $K/p_T$  
accounts for
the deflection of the track caused by the 
1.4T solenoidal magnetic field,
$\sigma_{K}$ accounts for multiple scattering
of the particle as it passes through the material of the
detector,
and $\sigma_{a}$ is present to account for any misalignment
between the COT and the detector to which the track is being
extrapolated.  The values of these constants are different for
each one of the five detectors to which the XTRP extrapolates
tracks.  The extrapolation window, which is the middle
term in the formula, allows for a 3-sigma multiple scattering
term combined in quadrature with a misalignment term.
See Table~\ref{Table:ExtrapVal} for typical values utilized
in extrapolation.


\begin{table}[htpb]
\bc
\caption{Typical values used in XTRP extrapolation.  The values of
$K$, $\sigma_{K}$, and $\sigma_{a}$ are constants that were
determined from data. The $p_T$ thresholds are set in the trigger
table and may change depending upon physics needs.  In the case of
the calorimeter trigger, electrons with a $2 {\rm GeV}/c$
threshold are identified separately from positrons with a $2 {\rm
GeV}/c$ threshold. In all other cases, the charge of the track is
not used to select leptons. } \label{Table:ExtrapVal} \ec \rm
\begin{tabular}{|l|l|l|l|l|}
\hline detector&$K$ ($^{\circ}/({\rm GeV}/c)$)&$\sigma_{K}$
($^{\circ}/({\rm GeV}/c)$)&$\sigma_{a}$($^{\circ}$) & $p_T$ thresholds
(${\rm GeV}/c$)\\ \hline
CMU&14.8&2.7&1.5 & 1.5, 4.0\\
CMX&13.36&5.39&1.5& 2.0, 8.0\\
IMU&0&6.11& 5& 5, 11.\\
CAL&8.72&1.22&1.5 & 2.0+, 2.0$-$, 4.0, 8.0\\
\hline
\end{tabular}
\end{table}

The values for the constants described above were initially 
set using Monte Carlo simulations of the data,
and the $\sigma_{a}$ term was conservatively
set to $5^{\circ}$ for all triggers to ensure
no events were lost.  After the initial running period of CDF-II, data was used
to refine the extrapolation
constants.  For the CMU and CMX
optimizations, $J/\psi$ data was used that came in on the CMU-CMU and CMU-CMX
triggers respectively.
The IMU trigger constants were examined using events with high momentum
muons.   For the calorimeter, generic tracks were 
used.  Because the calorimeter segmentation is coarse ($15^\circ$),
and a track traverses very little material between the COT and
the calorimeter, $\sigma_{K}$ is quite small.

In addition to providing a $\phi_{detector}$ window for
extrapolated tracks, the XTRP also implements $p_T$,
charge and short track selection criteria.
The values of these cuts can
change run-by-run and are set by the CDF-II 
trigger table.  Typical momentum cut values are
also summarized in Table~\ref{Table:ExtrapVal}.


To allow flexibility in the trigger definitions, the lookup
tables utilized in the Data Boards are generated dynamically
at the beginning of each run.  This 
is achieved by providing the relevant parameters
to the initialization code, which in turn generates the lookup
tables within the VME processor in the XTRP crate.  This provides
the additional advantage of transferring a small amount of
data to the crate as opposed to 
downloading the large lookup tables
over the network.
On initialization, the XTRP VME processor is provided with
the $p_T$, charge and short-track criteria from
the trigger table and the extrapolation constants from a
hardware database.  From this information, the code
generates 12 sets of extrapolation maps, one
for each of the RAMs in an XTRP wedge.  It then sends copies of
the sets of maps to each of the 24 wedges, so that all of the 288
RAMs has the appropriate look-up table.  It is only because the
detector is nearly azimuthally symmetric and each XTRP wedge is
identical that we need only 12 sets of maps;  if necessary, we
could generate 288 distinct lookups, one for each RAM.  Since
there is a 24-fold symmetry, to save time we have configured
the system so that each lookup is simultaneously written to
the 24 RAMs.

The Track Trigger lookups are generated in a manner analogous
to that described above.  For the Track Trigger, the trigger table
provides the parameters and bit assignments for each of
the triggers.  The lookups are generated within the VME CPU and
written to the lookup RAMs by VME block transfer.
\section{XTRP System Testing}
\label{sec:XtrpTesting}


The XTRP system is designed with an eye toward testing and diagnosing
system integrity.  Functionally, the system is implemented via synchronous lookup
rams and parallel pipelines stepping at 33ns.  The design allows for
VME readback of
data at each stage in the pipelines which make possible sophisticated diagnostic
and test procedures.  These include basic functionality tests, full bit-by-bit tests,
and full speed system tests.  Within the trigger system,
the XTRP is expected to provide
reliable trigger information at a rate of 7.5Mhz with negligible error rates.  This
section describes the system features and software used to verify the system integrity
and reliability, long before actual tracking data was available from the CDF-II detector.

All testing and diagnostic software was developed using CDFVME software framework.  This
package was developed to provide board and test stand software for CDF VME data acquisition
and trigger system.
It allows users to write java GUIs which communicate via ROBIN protocol with VxWorks C
routines 
running on the VME front-end 
processor \cite{vxworks,fision,robin}.  This enabled convenient development of
Java-based graphical user interfaces (GUIs) while maintaining good system I/O performance on the VME
front-end.

\subsection{Basic Functionality Testing}
\label{subsec:CoreFunc}

All diagnostic and system testing software is based upon VME transactions with the XTRP
system boards.  With this in mind, the testing of each board is predicated upon a
working VME interface.  The VME interface is handled by a dedicated FPGA on each board
which is loaded from a PROM at power-up.  Each FPGA design contains a bank of test
read/write
registers to verify basic VME interface functionality and debug potential board
level problems in the VME bus between the backplane and the VME FPGA.  A reliable VME
interface is required to load auxiliary FPGAs on the system boards and is 
the primary tool
to probe basic functionality.

As noted in Section~\ref{sec:XtrpDesign},
on-board RAMs play a large role in the operation of the XTRP system.
Before more sophisticated tests can be run, each RAM is exercised extensively to verify
reliable operation.  This consists of writing varied patterns and reading them back
successfully from each RAM over thousands of trials.

On the XTRP Data Boards, the output of each RAM is the input into a GTL whose output is ORed
onto a common bus.  This first set of GTLs perform the intra-wedge ORing and
 are  referred to as stage 0.  As there are 12
GTLs which are ORed together onto this bus, it can be difficult to diagnose the source
of the problem if a bit error occurs on the stage 0 output.  As part of the 
on-board
diagnostics, the input to each GTL is tied to the VME bus.  There are buffers which not
only allow readback of the input of the GTL from the RAM under normal operation but also
provide a means to bypass the RAM output and provide input to the GTL directly.  Using
these features, dedicated tests were developed to individually enable and test each bit
independently on both the input and output sides of the GTLs, allowing for quick
identification of any unreliable connections. Similar tests were developed and
utilized to check the input and outputs of the 
inter-wedge ORing stages 1 and 2.

At the heart of the XTRP system is the system clock generated by
the XTRP clock board. The clock functions in three basic modes -
VME, normal, and burst.  VME mode is used to generate the edges of
the clocks ($33\rm~ns$ \& $132\rm~ns$) via 
VME writes to a specified register on the
clock board.  This is only used for testing and
diagnostics.  In normal mode the clocks are free running and
driven from the input system clock (CDF Clock).  Finally there is
burst mode in which a programmed number of clocks are generated
after a given trigger event.  Each of these modes play an important
role in testing
and diagnosing the XTRP system operation, and are discussed
in detail below.

\subsection{Emulation}
\label{subsec:SimJava}

To test the boards a full bit-by-bit emulation of each of the
XTRP boards was developed.  The emulation was written in java and
provided classes for each board in the system to simulate all XTRP
system functions including Level-1 trigger pipelines, Track
Trigger, and Level-2 tracklists simultaneously.  At the macro
level, the emulation software is given track data input to each
of its Data Board objects along with corresponding clock inputs
and then provides all corresponding outputs as the data is clocked
through the system.  On a micro level, the emulation also
provides full data for each pipeline throughout the system,
from input to each RAM through each subsequent step of 
system. 
The emulation combined with the extensive VME
diagnostic readback in the system provides a means to
identify and diagnose board level problems in the entire
system.

\subsection{VME Clock Tests}
\label{subsec:Clocking}

\bfi[htbp] \bc
\includegraphics[width=14cm]{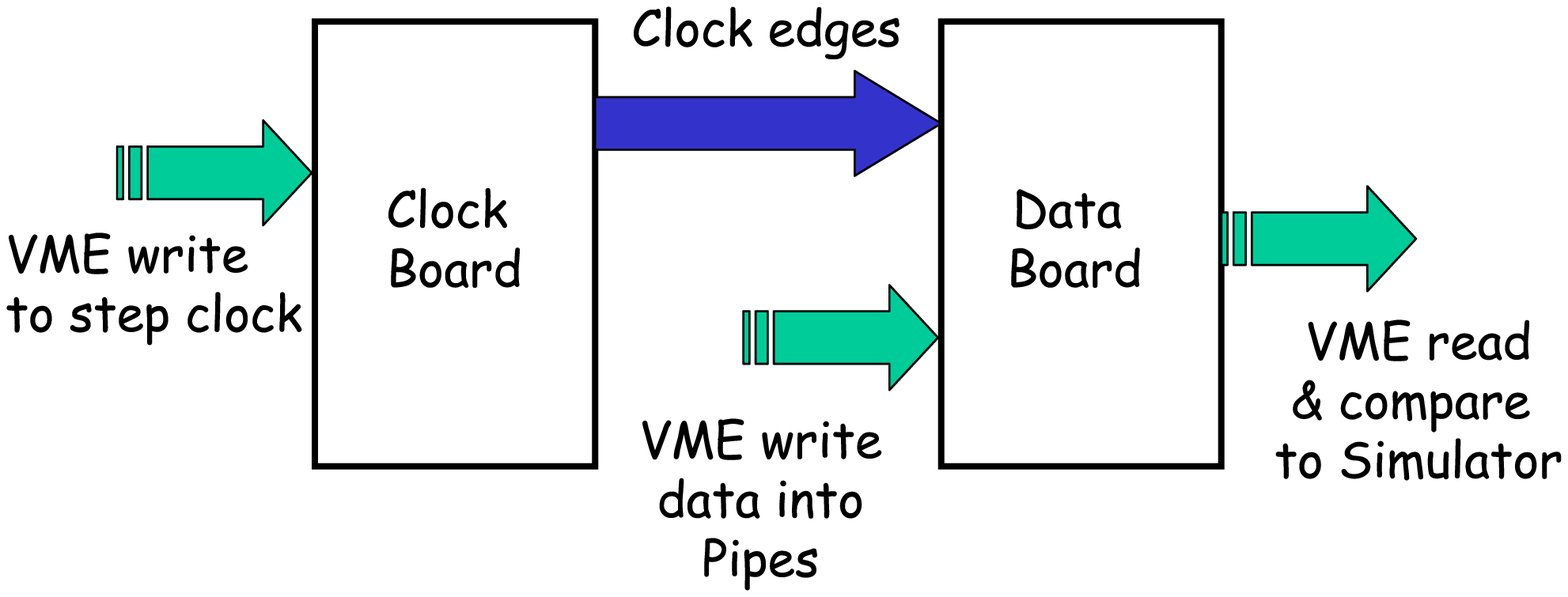}
\ec \caption{Block diagram of the VME test procedure where data is
manually inserted and stepped through the system under VME
control.} \label{fig:VMETest}
\efi

\bfi[htbp]
\bc
\includegraphics[width=14cm]{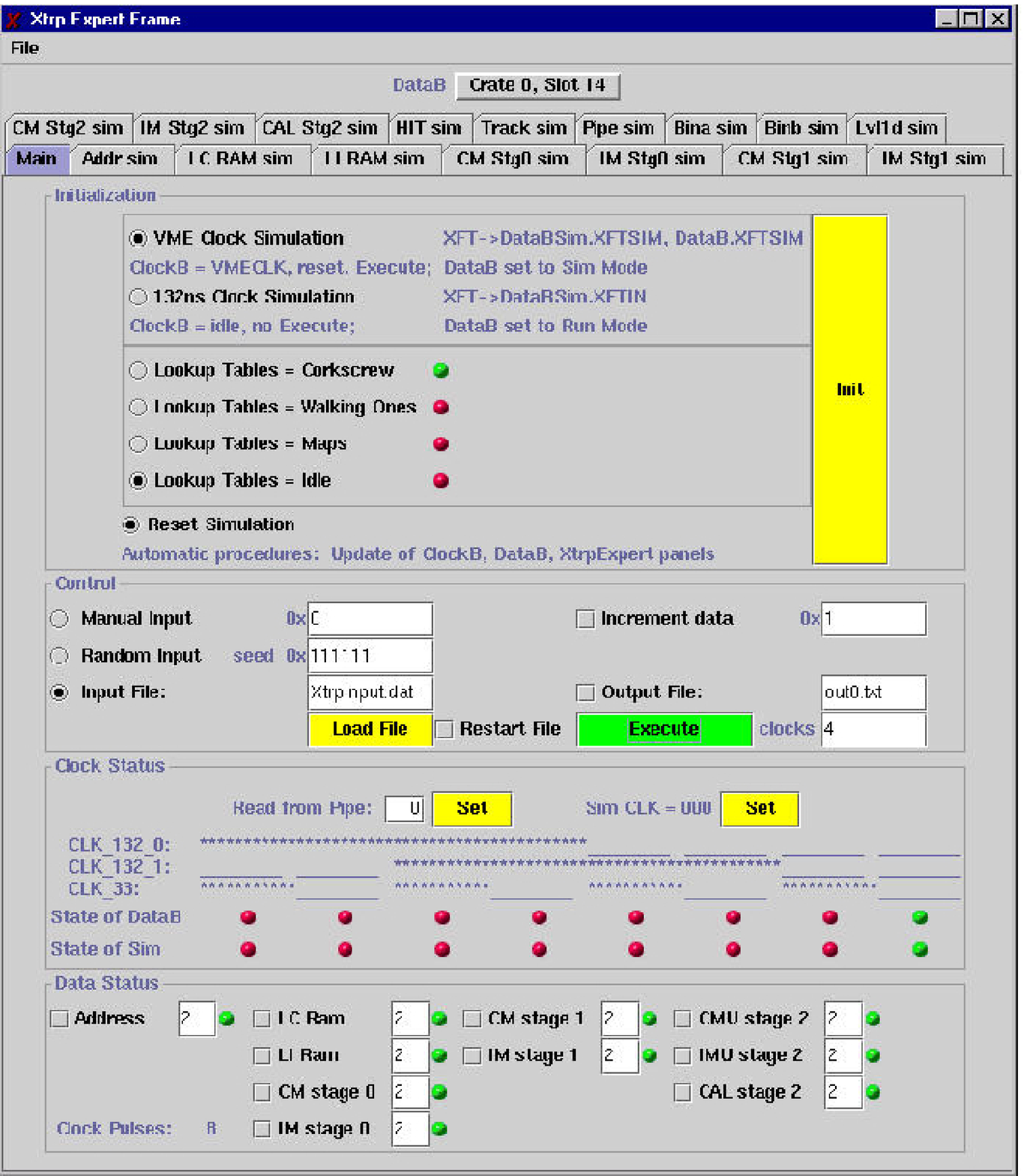}
\ec \caption{Main display for the ExpertPanel diagnostic
application.  Allows the user to control configuration and
setup while reporting on the status of hardware to emulation
comparisons.} \label{fig:expertPanel}
\efi

The most thorough system tests are performed by using the VME
interface to provide clock inputs to the XTRP system.  In this
scenario it is necessary to provide a method to input data into
the system which is also easily controlled from VME to synchronize
with the clocks.  This is accomplished by taking advantage of the
Pipe FPGAs on the Data Boards.  Under normal operation, these
FPGAs receive incoming data from the transition modules, but in
the VME test mode, they are switched to use input from internal
VME writable registers.  In this manner, it is possible to inject
data and step it through the system entirely under user control
via VME.  A block diagram of the VME test is shown in Figure
~\ref{fig:VMETest}.

A complete GUI interface referred to as the ``ExpertPanel'' was
developed to exercise the system under VME control.  It interfaces
not only with the XTRP hardware but also with the emulation
software so that it can compare actual data in the boards to
emulation expectations throughout the system.  Figure
~\ref{fig:expertPanel}, shows the main control display for the
ExpertPanel.  It provides control over system setup, input data,
and clocks as well as an overview of the data to emulation
comparison at each stage in the XTRP system.  Each stage also has
its own subpanel accessible via the tabbed panes at the top,
which provides full bit-by-bit comparisons of the data at that
stage in the board that can be used to pinpoint the exact location
of any discrepancies.  The drawback to this method is that it is
very slow and it would take an incredibly long time to step large
amounts of data through the system.

\bfi[htbp] \bc
\includegraphics[width=14cm]{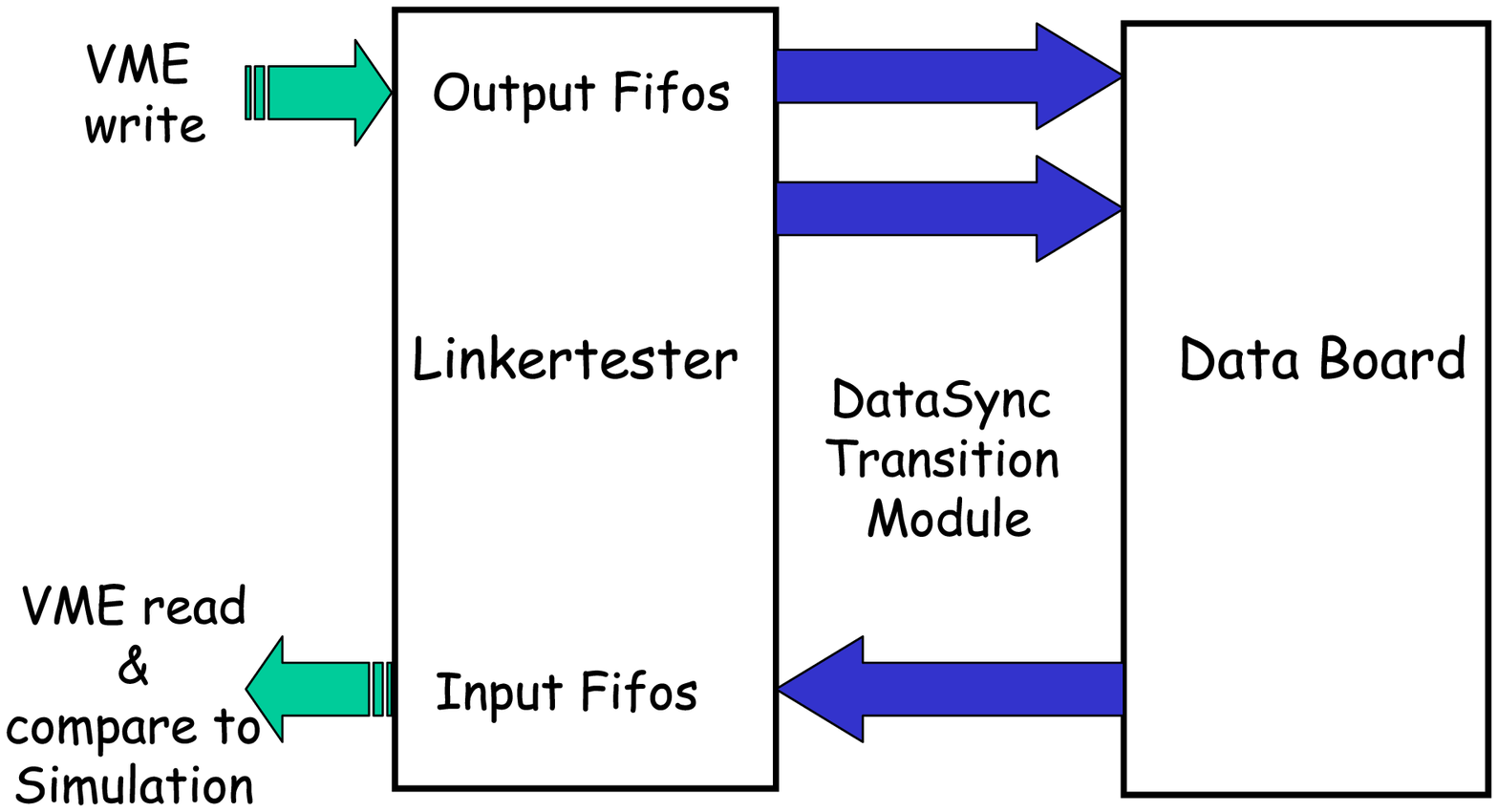}
\ec \caption{Block diagram of test to insert and readback data
from the system at full speed.} \label{fig:fullSpeedTest}
\efi

\subsection{Full Speed Testing}
\label{subsec:SpeedTest}

In order to increase the data bandwidth and test for timing
related issues, the ability to process data with the Clock/Control Board
and all system clocks operating at full speed was developed.  At
this speed data on the board is changing every $33\rm~ns$, so it is
impossible to attempt to see what is happening by VME diagnostics.
For full speed testing, data is input from an external source and
the board outputs are subsequently logged and examined as shown in
Figure~\ref{fig:fullSpeedTest}.

This required some additional test specific hardware which was
developed for testing the XFT system.
The LinkerTester, consists of a number of 
very large FIFOs tied to
input and output channels and synchronized by the CDF Clock and
programmable marker signals.  Using this board, the output FIFOs can be
configured to hold over 10k events for a single XTRP Data Board
and are capable of capturing all the subsequent Level-1 trigger output.

Using the LinkerTester board, it was possible to run large amounts of data into the
system at full speed and record the resulting trigger output which
could be readback and checked against the expected emulation results.  An application,
referred to as LoopTest, was developed to continuously exercise the system at full
speed and report any errors.  This also exercises the full input and output path on the
XTRP Data Boards which was not done by the VME tests.

LoopTest can run the system at full speed and determine if an error occurs in the output,
but it does not provide any diagnostics about the error other than the input data which
resulted in the error.  It is then possible to input the same data at VME speed with the
ExpertPanel to see if there is an error in the logic. If no error is observed at
VME speed, then the error can be checked using the burst mode
feature of the XTRP clock board.  
In this mode the system can be run at full speed for
an programmable number of clock cycles 
and then halted so that the ExpertPanel 
can be used to readback the state of the board
and compare to emulation expectations.

\subsection{Interface Testing}
\label{subsec:InterfaceTest}

The final phase was to verify that the XTRP system was able to interface with other
CDF-II trigger systems.  As shown in Figure~\ref{fig:trigger}, the
XTRP system receives its input from
the XFT continuously and provides continuous output to Level-1 muon trigger,
calorimeter trigger, and Level-1 global trigger.  It also provides output to L2 trigger
and SVT system when a L1 trigger accept is issued.

\bfi[tbp] \bc
\includegraphics[width=14cm]{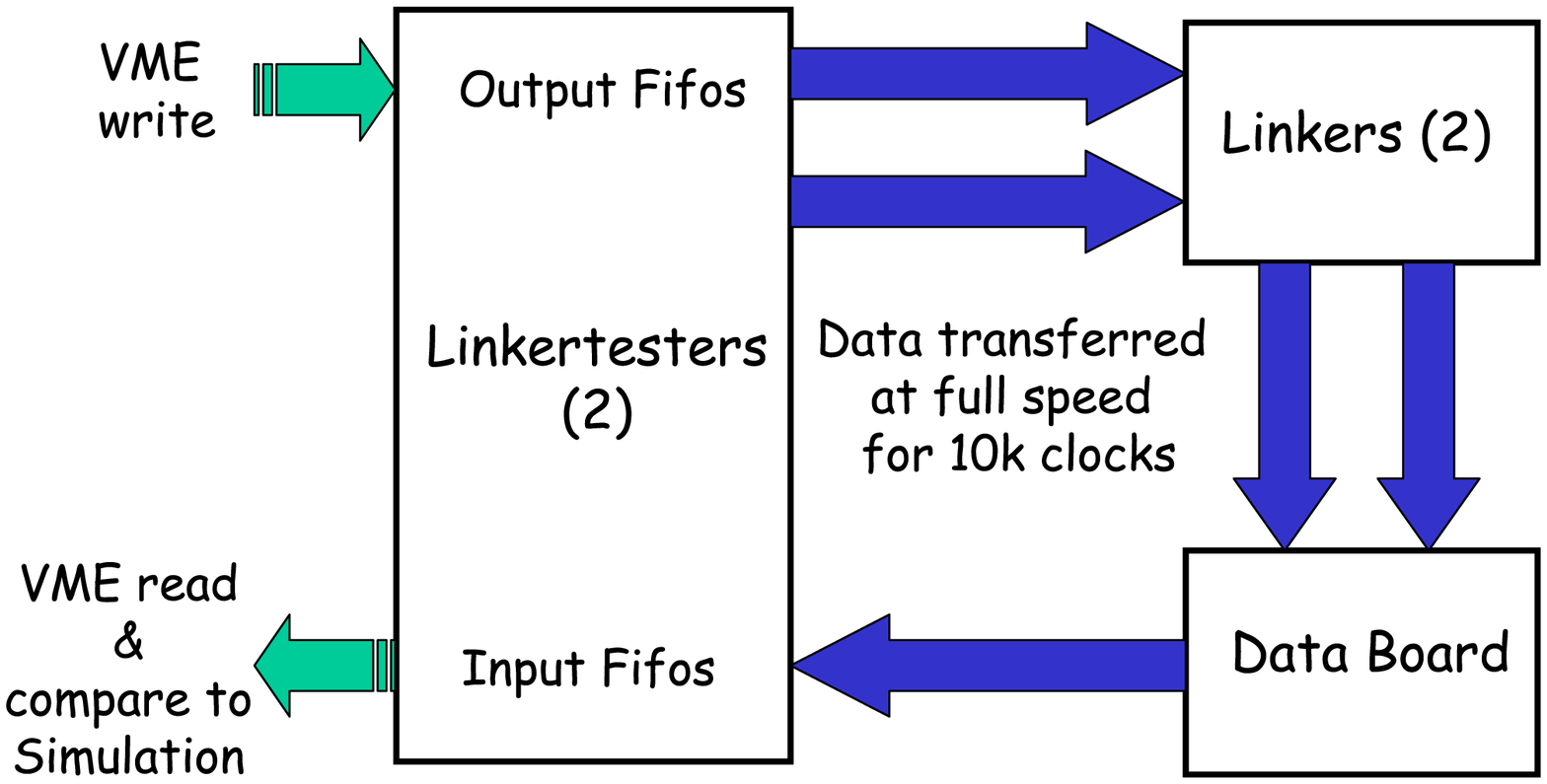}
\ec \caption{Block diagram of the setup used to test the data
interface between the XFT Linker and XTRP Data Boards.} \label{fig:linkerTest}
\efi

The most extensive interface testing was done between the XFT
Linker and the XTRP Data Boards.  This link transfers a total of
1152 bits every 33ns over a total of 24 cables from 24 different
Linkers into 12 XTRP Data Boards where they all must be latched at
the same time to keep the data synchronized.  This required
extensive testing to determine the optimal timing settings for the
system and insure reliable data transfer. The LinkerTester,
being originally designed to test the XFT Linker boards, was used to
perform these tests.  The optimal timing settings were determined
by attaching a logic analyzer to the incoming signals and
adjusting the XTRP system clock using the delays on the cdf clock
board.  Reliability tests were then performed by running large
volumes of data through both the Linker and XTRP using the 
LinkerTester as shown in Figure ~\ref{fig:linkerTest}.

The output interfaces were much simpler and were exercised during the initial testing
but did not get fully exercised until all the trigger systems were installed in the CDF-II system.  
This was done during commissioning of the CDF-II detector and electronics.  The XTRP
system became a valuable tool during this period.  A new design for the pipe FPGAs
on the Data Boards was developed which allowed for simulated track data to be input
and run through the system.  With beam time for testing at a premium, this proved an
invaluable tool not only for interface tests but for testing and commissioning track
based triggers throughout the system.

\subsection{System Timing}

As described above, the relative timing between the
XFT and XTRP systems was well measured in the
test stand.  At the time of installation of the trigger
system, the overall system timing was known
relatively well (to within $\pm 1$ clock cycle) based
upon a detailed accounting of measured latencies and
cable delays.  The overall system timing was then 
verified using colliding beam data, where events 
were read out on four consecutive clock cycles.  By
comparing the data for each subsystem in each of
the four events, the final timing of the system 
could be established and set.   This technique was
straightforward because of the detailed accounting
performed prior to operation.   Once the system was
fully timed-in, the timing has been stable and robust. 
\section{Operations}
\label{sec:Operations}

The XTRP system was installed and commissioned during
the Summer 2001.  It was fully functional for the
CDF-II physics run that began in January 2002.

\subsection{Maintenance and Reliability}
\label{subsec:Maintenance}

The reliability of the system has been quite high.  In 3.5
years of running, there have been two single-component failures
on Data Boards, and zero failures on any of the other boards
in the system.  In
both cases of Data Board component failure,  
the board was replaced while it was repaired.

System maintenance has been successful for three
reasons:
\begin{enumerate}
\item This digital system can be fully emulated to
   quickly identify problems.
\item The testing software described in the Section~\ref{sec:XtrpTesting}
  make it possible to quickly isolate the source/location
  of an error.
\item The XTRP system is located away from the beamline,
   so it can be accessed at any time
   without loss of colliding beams.
\end{enumerate}
In the following section, we describe the emulation and
monitoring tools that are utilized in this system.

\subsection{Monitoring: XTRPSim and XTRPMon}

\label{subsec:Monitoring}

As described in Section~\ref{sec:XtrpTesting}, we are able to
fully emulate the performance of this digital system.  We
have constructed a software package, known as XTRPSim,
which can use data from the detector (or data from
simulated events) to fully emulate the XTRP system.  This
digital emulation was utilized heavily during system
development and commissioning, and was then ported to the
CDF-II analysis environment so that it could be utilized
while the experiment is running.  

During data-taking, a random sampling of events are fed
to the ``monitoring stream'' at a rate of $1\rm Hz$.  These
events are fully emulated using the XTRPSim.  The results
of the emulation, along with the data, are then passed to
XTRPMon, which performs several checks on the data.  XTRPMon
performs bit-for-bit comparisons between XTRP data and the
XTRPSim emulation.  In
addition, many relevant pieces of trigger information are 
read out at different locations in the trigger chain.  These
can be compared to one another to check for data link integrity.
For example, the XFT track data is available from the XFT, 
XTRP, SVT and Level-2 trigger systems.  Checking that these
track lists are identical to one another is a powerful tool to
monitor the digital links between these systems.  The
$1 \rm Hz$ sampling of events is sufficient to identify 
problems without utilizing significant bandwidth.
In the control room, XTRPMon provides monitoring histograms
to the shift crew.

\bfi[tbp] \bc
\includegraphics[width=15cm]{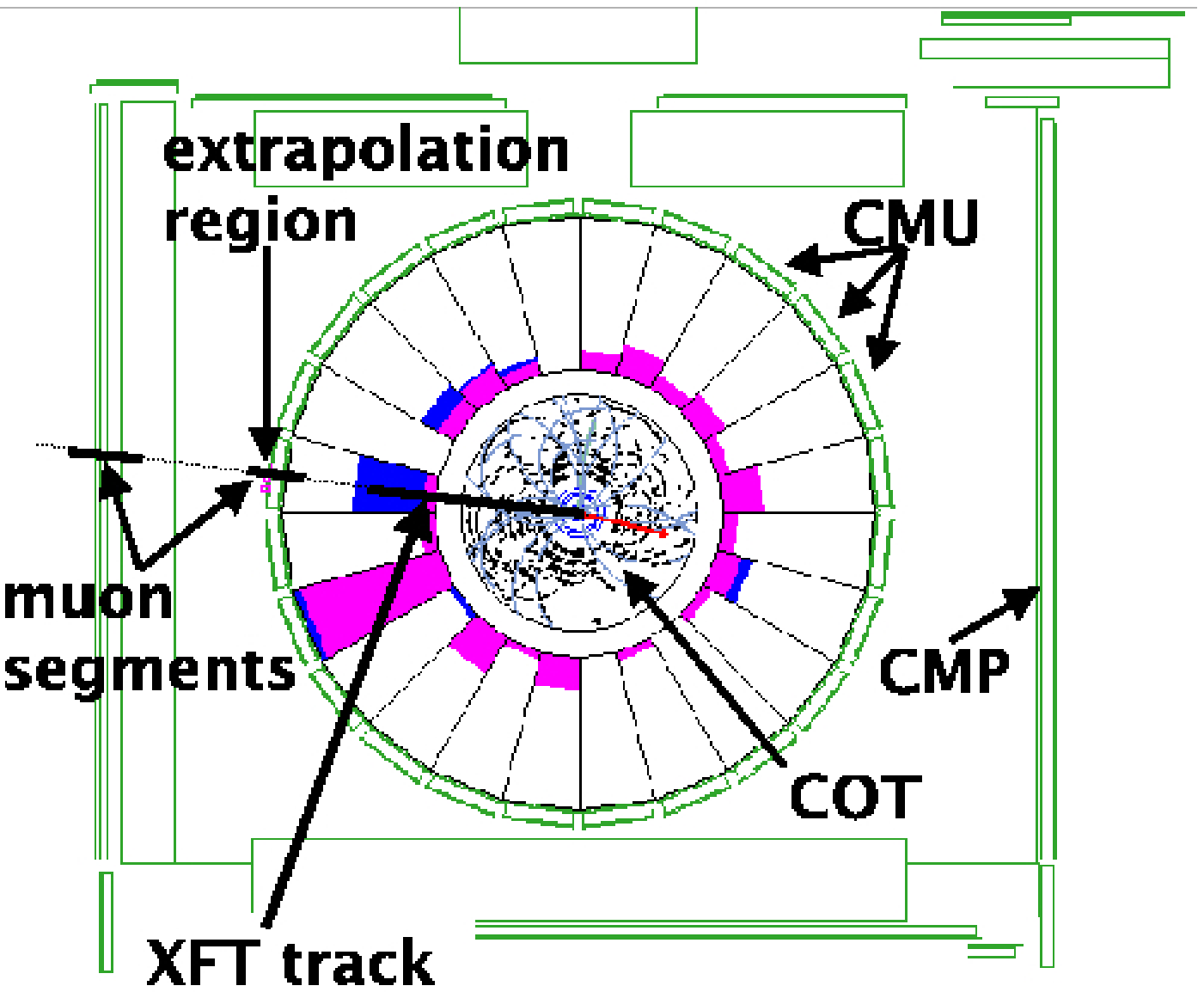}
\ec \caption{A colliding beam event taken with the CDF-II
detector.  The event was accepted via a high-$p_T$ 
muon trigger path.  The COT is shown at the center of
the figure.  There are several tracks observed with 
$p_T<1.5\, \rm GeV/{\it c}$ that are 
reconstructed offline, but below the momentum
threshold of the XFT.  The one high momentum XFT track
is identified pointing to the left in the figure.  
The central
calorimeter energy is shown just outside the COT, followed 
by the CMU, shown as 24 rectangular detectors.  The CMP
is shown as the outer 
``box" shaped detector.
The
XFT track was extrapolated to the muon system and matched with 
track segments in the CMU and CMP detectors.  The extrapolation
``window" is shown as three small squares just outside the
CMU detector.  This indicates that the XTRP required a CMU
stub within a $6.3^\circ$ extrapolation window that accounts
for curvature, misalignment and multiple scattering.
} \label{fig:event}
\efi

\subsection{Performance}
\label{subsec:perform}

The XTRP system is performing as specified.  Track matching
to muons with a $2.5^\circ$ granularity 
in the trigger allows for high
purity single and dimuon triggers.  Figure~\ref{fig:event}
shows a high momentum muon identified at the trigger level.
Track matching to electrons
with $15^\circ$ granularity (followed by tighter matching
at Level 2) allows for single and dielectron triggers.
The Track-Trigger allows for single and two-track triggers
based upon charged tracking information.

Overall, approximately $75\% $ of the CDF-II trigger bandwidth
utilized track-based triggers (electrons, muons and tracks)
by way of the XTRP system.    The flexibility of the look-up
based extrapolation and trigger maps have allowed for
improved and optimized trigger algorithms and matching
parameters.  The flexibility of the system firmware has also
permitted algorithmic improvements.

\subsection{Physics Impact}
\label{subsec:PhysImpact}

It is difficult to quantify the precise physics impact
of different components of an integrated trigger system, but
there are some aspects of the CDF-II physics program that
require the full performance of the XTRP system:
\begin{itemize}
\item low-momentum dimuon triggers.  We allow for muons with
$p_T>1.5\, {\rm GeV}/c$ for dimuon events, providing a large yield
of $J/\psi\rightarrow \mu^+ \mu^-$ decays.  This mode is important
for $B$ physics as well as detector calibration \cite{dimu}.
\item high-momentum single muon triggers \cite{mu}.   
We allow for single
muons with $p_T>4\, {\rm GeV}/c$.  The trigger rate is manageable
thanks to the precise matching that XTRP provides between the 
found tracks (XFT) and muon segments (L1 MUON).
For muons with $p_T=10\, {\rm GeV}/c$, the extrapolation
window in the trigger is approximately $1^\circ$, a precision
that has never been achieved at the trigger level.
\item single and di-electrons.  With four possible electron
thresholds, our standard thresholds are 
$p_T = 8\, {\rm GeV}/c$ for inclusive high momentum electrons, 
$p_T = 4\, {\rm GeV}/c$ for dilepton search triggers and semielectronic
$B$ decays, and $p_T = 2\, {\rm GeV}/c$ for $J/\psi\rightarrow e^+e^-$ \cite{ele}.
\item track-only triggers.  The CDF-II experiment has revolutionized
hadron collider heavy flavor physics (charm, bottom) with the
tracking trigger path that utilizes the Level 1 Track Trigger
followed by the Level 2 Silicon Vertex Tracker.  This has allowed
large samples of hadronic charm and bottom decays to be identified.
Hadronic heavy flavor physics was simply not accessible before this
system was implemented \cite{hf}.
\end{itemize}

The modularity of the CDF-II trigger system lends 
itself nicely
to upgrades.  We are currently commissioning
the Track Trigger-II
system, which will be a replacement to the 
Track Trigger system described in this document.

The design of the Track Trigger-II takes advantage
of our experience with the XTRP/Track Trigger
system and will be a direct
replacement to the existing Track Trigger system.
Improvements in SRAM 
technology, as well as faster, less expensive
FPGA technology permit a design of the 
Track Trigger-II that will 
be able to process more tracks per event than
the existing system.  This is important at higher
instantaneous 
Tevatron luminosity, because the track multiplicity
grows with the number of interactions per bunch
crossing.  In addition, by performing data
compression along with additional preprocessing,
the Track Trigger-II is able to select events
based upon the transverse mass of track-pairs.
The new system provides a significant improvement
in trigger performance, providing greater 
background rejection and better signal purity.
\section{Conclusion}
\label{sec:conclusion}
 
We have developed a track extrapolation system and distribution
system for the CDF-II trigger.  The system was designed with testing,
commissioning and monitoring in mind.  The XTRP system has been functioning
as part of the CDF-II trigger since the beginning of Tevatron Run~II,
and we anticipate it will remain an integral part of the trigger 
for the remainder of the run.  The XTRP system, and the CDF-II trigger
as a whole, are providing unprecedented data samples and 
access to physics channels never before observed.

We thank our colleagues on the CDF-II experiment, as well as the
Fermilab staff.  This work supported by Department of Energy,
Contract DE-FG02-91ER40677.




\end{document}